\documentstyle[preprint,aps,eqsecnum]{revtex}

\begin{document}
{\tighten
\preprint{\vbox{\hbox{UCSD/PTH 93--38}\hbox{CALT--68--1910}
\hbox{JHU--TIPAC--930029}}}

\title{QCD Corrections and the Endpoint of the Lepton Spectrum\\
in Semileptonic $B$ Decays}

\author{Adam F.~Falk\thanks{On leave from The Johns Hopkins University,
Baltimore, Maryland}, Elizabeth~Jenkins and Aneesh V.~Manohar}
\address{Department of Physics\\
University of California,~San Diego\\  La Jolla, California 92093}

\author{Mark B.~Wise}
\address{California Institute of Technology\\ Pasadena, California 91125}

\bigskip
\date{December 18, 1993}

\maketitle
\begin{abstract}
Recently, Neubert has suggested that a certain class of nonperturbative
corrections dominates the shape of the electron spectrum in the endpoint
region of semileptonic $B$ decay. Perturbative QCD corrections are
important in the endpoint region. We study the effects of these
corrections on Neubert's proposal. The connection between the endpoint of
the electron spectrum in semileptonic $B$ decay and the photon spectrum
in $b\rightarrow s\gamma$ is outlined.
\end{abstract}

}

\section{Introduction}

The electron energy spectrum near its endpoint in semileptonic $B$ meson decay
arises from $b\rightarrow u$ transitions and provides one method for the
extraction of the Kobayashi-Maskawa mixing angle $V_{ub}$ from experiment.
The spectrum must be known accurately within a few hundred MeV of its endpoint,
since it is only in this region that the large background due to the dominant
$b\rightarrow c$ weak transition is kinematically forbidden.  Thus, the
separation of the rare $b \rightarrow u$ decay from the inclusive spectrum
relies upon a theoretical understanding of the shape of the spectrum in this
small region.  Unfortunately, it is precisely this region which is the least
well understood theoretically.

The endpoint region of inclusive semileptonic $B$ decay has been studied
extensively.  The first approaches relied on QCD models.  Grinstein et
al.~\cite{ISGW} used a constituent quark model to sum over exclusive charmless
final states in this region, assuming that the spectrum is dominated by a few
low-lying resonances.  Altarelli et al.~\cite{ACCMM} computed the spectrum in
the free $b$ quark decay model, augmented by the inclusion of a model of the
Fermi motion of the $b$ quark in the $B$ meson.  More recently, a QCD-based
approach has been formulated in the context of the heavy quark effective theory
(HQET).  Using an operator product expansion (OPE) and the HQET,  Chay, Georgi
and Grinstein~\cite{CGG} have shown that the free $b$ quark decay model
describes inclusive semileptonic $B$ decay to leading and first subleading
order in a systematic expansion in $1/m_b$, where $m_b$ is the $b$ quark mass
of the HQET. The first non-vanishing corrections to the free quark decay
result are of order
$1/m_b^2$, and have now been computed~\cite{MW,Bigietc}.  These corrections
arise from higher order terms in the OPE whose matrix elements contain
information about the state of the $b$ quark inside the hadron.

At leading order, the electron spectrum is governed by quark kinematics with an
endpoint at $E_e=m_b/2$, rather than at the physical endpoint $M_B/2$ which is
determined by the $B$ meson mass $M_B$.  The higher order terms in the $1/m_b$
expansion produce corrections to the free quark decay spectrum, causing it to
``leak'' beyond the free quark endpoint. Understanding this process is crucial
for extracting $V_{ub}$, since the difference $(M_B-m_b)/2$ is expected to be
several hundred MeV, and is comparable to the 330~MeV energy difference between
the $b\rightarrow u$ and $b\rightarrow c$ endpoints. Recently, Neubert has
shown that the most singular terms in the $1/m_b$ expansion can be used to
define a ``shape function'' of the spectrum, which is determined by a certain
set of nonperturbative matrix elements and is
model-independent~\cite{Neubert}.  This shape function describes the electron
energy spectrum beyond the kinematic endpoint of the free quark decay
(neglecting QCD radiative corrections). In this paper we examine the influence
of perturbative QCD corrections on the endpoint region. These corrections are
particularly important due to the presence of a Sudakov double-logarithmic
suppression of the free quark decay rate at the endpoint.

This paper is organized as follows.  In Section 2, we review the operator
product expansion analysis of the differential decay width for the endpoint
region, neglecting perturbative QCD radiative corrections. The summation of the
leading nonperturbative singularities to the shape of the endpoint spectrum is
presented.  We show that this summation can be obtained from the free $b$ quark
decay result by suitably averaging the free quark decay result over the
residual momentum of the $b$ quark inside the $B$ meson.  Since the leading
nonperturbative corrections can be generated by this procedure, radiative
corrections can be included by computing radiative corrections to the free
quark decay result and then averaging over the residual momentum of the $b$
quark. In Section 3, we consider the radiative corrections to free quark decay
and show how they modify the shape of the endpoint of the electron spectrum.
Numerical results and conclusions are presented in Section~4.

\section{Leading Nonperturbative Singularities}

The inclusive differential decay distribution for $B\rightarrow
X_{u,c}\,e\,\overline\nu$ is determined by the imaginary part of the
time-ordered product of two weak currents,
\begin{equation}\label{Tmunu}
     T^{\mu\nu}\equiv-i\int d^4x\,e^{-iq\cdot x}\langle B | \, T \{ J^{ \mu
     \dagger} (x),J^\nu(0)\}\,|B\rangle\,,
\end{equation}
where $J^\mu=\overline q\gamma^\mu(1-\gamma^5)b$ and $q=u,c$.  The time-ordered
product may be expanded in inverse powers of the $b$ quark mass using an
operator product expansion~\cite{CGG}, and in powers of $\alpha_s(m_b)$.  In
this section we will concentrate on the $1/m_b$ expansion. From the operator
product expansion of the hadronic tensor, one obtains an expression for the
inclusive electron energy spectrum, $d\Gamma/dy$, where $y$ is the rescaled
electron energy, $y=2E_e/m_b$. The leading term in the $1/m_b$ expansion
produces the result of the free quark decay model, in which the inclusive
semileptonic decay rate is given by the decay of a free, on-shell $b$ quark.
The endpoint of the electron spectrum is at $y=1$. The subleading terms
represent corrections to free quark decay, in which certain features of the
motion of the $b$ quark inside the $B$ meson are taken into account.  The
expansion is in powers of
\begin{equation}
     \epsilon=\Lambda/m_b\,,
\end{equation}
where $\Lambda$ is a scale typical of the strong interactions of QCD, perhaps
300 to 500 MeV.

Neglecting perturbative $\alpha_s(m_b)$ corrections, the electron energy
spectrum for
$B\rightarrow X_u\,e\,\overline\nu$ decay is given by~\cite{MW,Bigietc}
\begin{eqnarray}\label{leading}
    {1\over\Gamma_0}{d\Gamma\over dy}&=&\left\{2(3-2y)y^2+4(3-y)y^2E_b
    -{4y^2(9+2y)\over3}K_b-{4y^2(15+2y)\over3}G_b\right\}\theta(1-y)
    \nonumber \\
    &&+\left\{2E_b-{4\over3}K_b+{16\over3}G_b\right\}\delta(1-y)
    +{2\over3}K_b\delta'(1-y)\,,
\end{eqnarray}
up to corrections of order $\epsilon^3$, where $\Gamma_0$ is the free quark
decay width
\begin{equation}\label{Gammazero}
     \Gamma_0=\left|V_{ub}\right|^2\,{G_F^2 m_b^5\over 192\pi^3}\,,
\end{equation}
and $\theta(x)$ is 1 if $x>0$ and zero otherwise.\footnote{Eq.~(\ref{leading})
holds for massless leptons. Lepton mass effects may be included \cite{tau},
but they do not change the behavior of the endpoint spectrum in an important
way.}  $E_b$, $K_b$ and $G_b$ are hadronic matrix elements of order
$\epsilon^2$, defined by
\begin{eqnarray}
     E_b &=& G_b+K_b\,,\cr
     K_b &=& \left\langle B(v)\right| \bar b_v\, {D^2\over 2 m_b^2}\, b_v
     \left|B(v)\right\rangle\,,\cr
     G_b &=& \left\langle B(v)\right| \bar b_v\,
     g{\sigma_{\alpha\beta}G^{\alpha\beta}\over 4 m_b^2}
     b_v\,\left|B(v)\right\rangle\,,
\end{eqnarray}
where $b_v$ is the $b$ quark field in the HQET.  The factor of $\theta(1-y)$ in
the first term is required because the tree level decay distribution does not
vanish at the boundary of the Dalitz plot. The $\delta(1-y)$ and $\delta'(1-y)$
singularities arise because some higher order terms in the $1/m_b$ expansion
have the form of derivatives with respect to $y$ of lower order terms. Since
the free quark decay distribution does not vanish at the endpoint, this
generates singular terms in the decay spectrum. These singularities imply that
the $1/m_b$ expansion breaks down at $y=1$.

Eq.~(\ref{leading}) is the decay spectrum including all corrections of order
$1/m_b^2$.  To all orders in $1/m_b$, the decay spectrum $d\Gamma/dy$ obtained
from the OPE at zeroth order in $\alpha_s$ has the structure
\begin{eqnarray}\label{generaldeltas}
     {1\over\Gamma_0}{d\Gamma\over dy} &=&
     \theta(1-y)\left(\epsilon^0+0\,\epsilon+\epsilon^2+\cdots\right)
     +\delta(1-y)\left(0\,\epsilon+\epsilon^2+\cdots\right)
     +\delta'(1-y)\left(\epsilon^2+\epsilon^3+\cdots\right)\nonumber\\
     &&+\cdots+\delta^{(n)}(1-y)\left(\epsilon^{n+1}+\epsilon^{n+2}
     +\cdots\right)+\cdots\,,
\end{eqnarray}
where $\epsilon^n$ denotes a term of that order, which may include a smooth
function of $y$. It is a nontrivial prediction of the heavy quark effective
theory that the terms proportional to $\epsilon$ in this expansion vanish
\cite{CGG}, as is evident in eq.~(\ref{leading}).  Although the theoretical
expression for $d\Gamma/dy$ is singular at the endpoint $y=1$, the total
semileptonic width is not. The contribution to the total rate of a term
$\epsilon^m\delta^{(n)}(1-y)$ is of order $\epsilon^m$, so the semileptonic
width has a well-behaved expansion in powers of $1/m_b$,
\begin{equation}
     \Gamma = \Gamma_0\left(1+0 \epsilon+ \epsilon^2 +\epsilon^3+\ldots\right),
\end{equation}
where the term proportional to $\epsilon$ vanishes.

The semileptonic decay width for $b\rightarrow u$ is difficult to measure
because of background contamination from the dominant $b\rightarrow c$
semileptonic decays. It is therefore important to be able to compute the
semileptonic decay rate for $b\rightarrow u$ transitions near the endpoint
$y=1$, since the kinematic endpoint of the $b\rightarrow c$ spectrum is below
the $b\rightarrow u$ endpoint. One way to calculate the endpoint spectrum is to
weight the differential distribution $d\Gamma/dy$ by a normalized
function of width $\sigma$ around $y=1$. We will refer to this procedure as
``smearing.'' Most of the details of the smearing procedure are unimportant;
the only quantity of relevance is the width $\sigma$ of the smearing region. A
physically meaningful result can be obtained by smearing over a large enough
region in $y$ such that the singular corrections to $d\Gamma/dy$ are small. In
ref.~\cite{MW}, it was shown that the singular corrections are small if
the smearing width is chosen so that $\sigma \gg \epsilon$. We will now
show that by summing the leading singularities, one can choose $\sigma$
of order $\epsilon$.

The singular distribution $\epsilon^m\delta^{(n)}(1-y)$ (where $m>n$) smeared
over a region of width $\sigma$ gives a contribution of order $\epsilon^m/
\sigma^{n+1}$ to $d\Gamma/dy$. If the width $\sigma$ of the smearing region
is of order $\epsilon^p$, the generic term $\epsilon^m\delta^{(n)}(1-y)$ yields
a contribution of order $\epsilon^{m-(n+1)p}$. Since $m>n$, this shows that the
$1/m_b$ expansion for the spectrum breaks down unless $p\le 1$, i.e. the
smearing region cannot be made narrower than of order $\epsilon$. If $p>1$, the
$1/m_b$ expansion breaks down because it is dominated by an infinite number of
terms at large values of $n$. This divergence is not associated with the
failure of the OPE due to the presence of resonances with masses of
order the QCD scale~\cite{Isgur}.   The region in which such resonances
dominate the final state is of width $\epsilon^2$, while the expansion
breaks down upon smearing over {\it any\/} region of size
$\epsilon^{1+\delta}$, where $\delta>0$.

If the smearing region is chosen to be of order $\epsilon$, the form of the
expansion (\ref{generaldeltas}) shows that the  leading terms of the form
$\theta(1-y)$ and $\epsilon^{n+1}\delta^{(n)}(1-y)$ all contribute at order
unity to $d\Gamma/dy$, all terms of the form $\epsilon^{n+2}\delta^{(n)}(1-y)$
contribute at order $\epsilon$, etc. Thus one can, in principle, obtain the
decay spectrum smeared over a width of order $\epsilon$ if one can sum the
leading singularities in eq.~(\ref{generaldeltas}). The sum of the leading
singularities produces a distribution $d\Gamma/dy$ of width $\epsilon$, and
with a height of the same magnitude as the free quark decay distribution for
$d\Gamma/dy$, i.e. with a height of order one. The subleading singularities
produce a distribution which is also of width $\epsilon$, but has a height of
order $\epsilon$ times the distribution obtained by summing the leading
singularities. The decay distribution $d\Gamma/dy$ cannot be obtained with a
resolution finer than $\epsilon$ without summing all the subleading
singularities.

Neubert has shown that the series of leading singularities
\begin{equation}\label{leadingseries}
     {1\over\Gamma_0}{d\Gamma\over dy}=A_0
     \,\theta(1-y)+0\,\epsilon\,\delta(1-y)
     +A_2\,\epsilon^2\,\delta'(1-y)+\cdots
\end{equation}
may be resummed into a ``shape function'', which describes the behavior of the
theoretical spectrum in the region beyond the free quark decay endpoint at
$y=1$ \cite{Neubert}.   These terms arise in a particularly simple way in the
OPE, because they come only from the expansion of the quark propagator which
connects the two currents.  The shape function has a width of order $\epsilon$
and height of order one.

The series of leading singularities (\ref{leadingseries}) can be obtained by
averaging the free quark decay result over the residual momentum of the $b$
quark in the $B$ meson \cite{MW}.  This simple procedure is important since it
will also enable us to obtain the leading nonperturbative singularities for the
radiative corrections by only calculating radiative corrections to free quark
decay.

The differential decay distribution is obtained from the tensor $T^{\mu \nu}$
defined in eq.~(\ref{Tmunu}). This tensor is a function of the momentum
transfer to the leptons, $q$, and the velocity of the $B$ meson, $v$.  The
differential decay distribution is proportional to the hadronic tensor
contracted with the lepton tensor $L_{\mu\nu}$, which depends on the electron
and neutrino momenta, $k_e$ and $k_\nu$:
\begin{equation}\label{dgform}
     {d\Gamma\over dx\, dy\, d\hat q^2} \propto W^{\mu\nu}L_{\mu\nu}\,,
\end{equation}
where $W^{\mu\nu}$ is the discontinuity of $T^{\mu\nu}$ across the physical
cut, $W^{\mu\nu}=-{\rm Im}\,T^{\mu\nu}/\pi$. The constant of proportionality
in eq.~(\ref{dgform}) involves $G_F^2$ and the mixing angle
$\left|V_{ub}\right|^2$. The dimensionless variables $x$, $y$ and $\hat q^2$
are defined by
\begin{equation}\label{xyq}
     x = {2 k_\nu\cdot v\over m_b}\,,\qquad y = {2 k_e\cdot v\over
     m_b}\,,\qquad \hat q^2 = {q^2\over m_b^2}\,.
\end{equation}
The lowest order (in $1/m_b$) decay distribution $d \Gamma_{\rm free}/ dx\,
dy\, d\hat q^2$ is the decay distribution for a free on-shell $b$ quark with
mass $m_b$ and the same velocity $v$ as the $B$ meson. However, the $b$ quark
in the $B$ meson is off-shell with a distribution of residual momentum $k$. The
off-shell $b$ quark, with momentum $m_bv+k$, may be viewed as an on-shell
quark with mass $m_b'$ and velocity $v'$, where $m_b'v'=m_b v+k$. The decay
rate for such a quark is obtained by evaluating the lowest order expression
for $d\Gamma_{\rm free}/dy$ in the rest frame of the moving quark, and then
boosting back to the rest frame of the $B$ meson,
\begin{equation}\label{Gprime}
   d\Gamma = {1\over v\cdot v'} \, d\Gamma_{\rm free}(x',y',\hat
   q^{\prime 2},m_b')\,.
\end{equation}
Note that all scaled quantities depend implicitly on $m_b$, and hence must be
primed.  We now replace $m_b'v'\rightarrow m_bv+k$ and average over the
residual momentum $k^\mu$.  Expanding in $k^\mu/m_b$, we obtain a series of
the form \cite{MW}
\begin{equation}\label{derivatives}
     \left\langle d\Gamma\right\rangle =
     \left\langle {\left[1 + 2v\cdot k/m_b + k^2/m_b^2\right]^{1/2}\over
     1 + v\cdot k/m_b}\left[1 + {k^{\mu_1}}
     {\partial\over\partial m_bv^{\mu_1}}
     +{1\over2} {k^{\mu_1}k^{\mu_2}}{\partial\over\partial m_b
     v^{\mu_1}} {\partial\over\partial m_b
        v^{\mu_2}}+...\right]d\Gamma_{\rm free}
     \right\rangle\,,
\end{equation}
where $\langle\cdot\rangle$ denotes an average with respect to the
distribution of the momentum $k$ of the $b$ quark in the $B$ meson. The
derivatives with respect to $m_b v^{\mu}$ can be  rewritten as derivatives
with respect to $x$, $y$ and $\hat q^2$ using the chain  rule. Terms with $n$
derivatives with respect to $m_b v^\mu$ in eq.~(\ref{derivatives}) turn into
terms with $n_x$, $n_y$ and $n_q$ derivatives with respect to $x$, $y$ and
$\hat q^2$ respectively, where $n_x+n_y+n_q\le n$. The expansion of
$d\Gamma/dy$ is then obtained by integrating the expansion of
$d\Gamma/dx\,dy\,d\hat q^2$ with respect to $x$ and $\hat q^2$. The explicit
computations to order $1/m_b^2$ are  given in ref.~\cite{MW}.

In this paper, we are interested in summing the most singular terms in
$d\Gamma/dy$ near $y=1$ to all orders in $1/m_b$. These terms are found by
retaining the terms in eq.~(\ref{derivatives}) with the maximum number of
$y$-derivatives at each order in $1/m_b$. This corresponds to only retaining
the $\partial^n/\partial y^n$ term in $\partial^n/\partial m_bv^{\mu_1}
\ldots \partial m_bv^{\mu_n}$ in eq.~(\ref{derivatives}) and ignoring the
prefactor $\left[1 + 2v\cdot k/m_b + k^2/m_b^2\right]^{1/2}/ \left[ 1 + v\cdot
k/m_b\right]$. Terms with derivatives with respect to $x$ or $\hat q^2$ do not
generate derivatives with respect to $y$ on integration over $x$ and $\hat
q^2$, and are less singular than the terms we have retained. The most singular
terms are thus obtained using
\begin{equation}\label{chainrule}
     \left({\partial \over\partial m_b v^\mu}\right)^n\rightarrow
     \left({\partial y\over\partial m_b v^\mu}{\partial \over\partial
     y}\right)^n\rightarrow\left( {2\over m_b}(\hat k_{e\mu}-yv_\mu){\partial
     \over \partial y}\right)^n
     \mathrel{\mathop{\longrightarrow}^{y=1}}\left({2\over m_b}(\hat k_e
     - v)_\mu
     {\partial \over \partial y}\right)^n\,,
\end{equation}
which gives the leading singularities,
\begin{eqnarray}
     {d\Gamma\over dy} &=&  {d\Gamma_{\rm free}\over dy}
     +\langle k^{\mu_1}\rangle \left({2\over m_b}\right)(\hat k_e - v)
     _{\mu_1}
     {\partial\over \partial y} \left({d\Gamma_{\rm free}\over dy}
        \right) \nonumber\\
     &&+\cdots+{1\over n!}\langle k^{\mu_1}\cdots k^{\mu_n}\rangle
     \left({2\over m_b}\right)^n(\hat k_e - v)_{\mu_1}\cdots(\hat k_e -
                               v)_{\mu_n}
     {\partial^n\over \partial y^n}\left( {d\Gamma_{\rm free}\over dy}
     \right)+\cdots\nonumber\\
     \label{derivseries2}
     &=&\sum_{n=0}^\infty {2^n\over m_b^{\,n}\ n!}(\hat k_e - v)_{\mu_1}
     \cdots(\hat k_e - v)_{\mu_n} \langle k^{\mu_1}\cdots
     k^{\mu_n}\rangle {\partial^n\over \partial y^n}
     \left({d\Gamma_{\rm free}\over dy}\right)\,,
\end{eqnarray}
where $\hat k_e = k_e/m_b$.  Eq.~(\ref{derivseries2}) sums the leading
nonperturbative corrections in the endpoint region, provided one interprets
the residual momentum $k$ in eq.~(\ref{dgform}) as the operator $iD$ and the
average as the expectation value of the resulting operator in the $B$-meson
state. There is no operator ordering ambiguity for the leading singularity in
this identification, because $D^{\mu_1} \cdots D^{\mu_n}$ is contracted with
the completely symmetric tensor $(\hat k_e - v)^{\mu_1}\cdots (\hat k_e -
v)^{\mu_n}$, and so the commutator $\left[D^\mu, D^\nu\right]$ does not
contribute.  Only the part of the matrix element $\langle  B
{(v)}|iD^{\mu_1}\cdots iD^{\mu_n} |B {(v)}\rangle$ proportional to the tensor
structure $v^{\mu_1}\cdots v^{\mu_n}$ contributes to the most singular terms,
since $(\hat k_e - v)^2$ vanishes at $y = 1$ \cite{Neubert}.  Neglecting
perturbative $\alpha_s (m_b)$ radiative corrections, the most singular terms
in eq.~(\ref{derivseries2}) are $\delta$-functions and their derivatives,
which arise from differentiating the factor of $\theta (1-y)$ in
$d\Gamma_{\rm free}/dy$. Dropping the $n=0$ term in eq.~(\ref{derivseries2})
and allowing the derivatives to act only on the $\theta$-function gives
Neubert's shape function
\begin{equation}\label{shapefunction}
     S(y)=\sum_{n=1}^\infty {2^n\over m_b^{\,n}\ n!}(\hat k_e - v)_{\mu_1}
     \cdots(\hat k_e - v)_{\mu_n}\ \langle B(v)|\, iD^{\mu_1}\cdots
     iD^{\mu_n}\,|B(v)\rangle\ {\partial^n\over \partial y^n}\,\theta(1-y)\,.
\end{equation}

This procedure for averaging over residual momentum produces the same result
for the leading singularities as the operator product expansion. As discussed
in ref.~\cite{MW}, one can use reparameterization invariance~\cite{Luke} to
show that averaging over residual momentum gives the same answer as the OPE,
provided one neglects the commutator $\left[D^\mu, D^\nu\right]$ and higher
dimension operators involving light quark fields. The commutator and higher
dimension operators do not contribute to the most singular terms, and so
averaging over residual momentum will be adequate for this discussion.

It is simple to understand how this averaging procedure generates a shape
function which extends beyond the free quark decay endpoint.  If the energy of
the $b$ quark is allowed to fluctuate from its on-shell value, occasionally it
will have an energy larger than its free value $m_b$.  This fluctuation
corresponds to a situation in which the quark has temporarily absorbed some
energy from the light degrees of freedom in the $B$ meson; if it decays weakly
at this moment, then an energy $E_e>m_b/2$ may be given to the electron.

\section{Radiative Corrections}

The advantage of the averaging procedure for obtaining the leading
nonperturbative singularities as $y\rightarrow1$ is that it generalizes
straightforwardly to the case when radiative corrections are included. The
averaging procedure applied to the free quark decay distribution including
radiative corrections yields the leading nonperturbative singularities
including radiative corrections.

The one-loop QCD contribution to the free quark decay process, including both
virtual gluons and real gluon emission, has been computed \cite{Alietc}.  The
corrected electron spectrum takes the form
\begin{equation}\label{radiative}
     {d\Gamma_{\rm free}\over dy}={d\Gamma_0\over dy}\left[
     1-{2\alpha_s\over3\pi}G(y,\hat m_q)
     +O(\alpha_s^2)\right]\,,
\end{equation}
where $\Gamma_0$ is the tree level free quark decay rate.
Perturbative QCD corrections do not extend the electron spectrum beyond the
free quark decay endpoint $y=1$. This can only occur because of the
nonperturbative $1/m_b$ corrections discussed in the preceding section. In the
interesting case $\hat m_q=0$ relevant to the transition $B\rightarrow
X_u\,e\,\overline\nu$, $G(y,0)$ is given
by~\cite{Alietc,JK}\footnote{The expression for $G(y)$ is taken from
ref.~\cite{JK}.}
\begin{equation}\label{Gdef}
     G(y,0)=G(y)=\ln^2(1-y)+{31\over6}\ln(1-y)+\pi^2+{5\over4}+
     (\text{vanishing as } y\rightarrow 1)\,.
\end{equation}
The leading
singularity at each order in perturbation theory is proportional to
$\alpha_s^n\ln^{2n}(1-y)$. These singularities lead to a breakdown of the
perturbative QCD expansion near the endpoint $y=1$, unless they can be summed.
The double logarithms have been shown to exponentiate \cite{Sudakov}, yielding
an expression which formally has the structure
\begin{equation}\label{exponentiated}
     {d\Gamma_{\rm free}\over dy}=R(y){d\Gamma_0\over dy}\,,
\end{equation}
where
\begin{equation}\label{Rdef}
     R(y)=\exp\left\{-{2\alpha_s\over3\pi}\ln^2(1-y)
     \right\}\,.
\end{equation}
This is the form for the decay spectrum used by Altarelli et al.~\cite{ACCMM}.
The Sudakov form factor $R(y)$ causes the electron spectrum to vanish
at the free quark endpoint $y=1$.

The contribution to the endpoint shape of the electron energy spectrum coming
from the exponentiated double-logarithm in $R(y)$ is a calculable effect. One
might hope that once this leading radiative correction has been accounted for,
it would be consistent to include the leading higher dimension operators using
eq.~(\ref{derivseries2}) and neglect all subleading radiative corrections.
However, we find that for very large $m_b$ this is not the case; the
perturbative expansion is so poorly behaved at large orders in $\alpha_s(m_b)$
that it is necessary to sum an infinite number of infinite series before
including nonperturbative effects with eq.~(\ref{derivseries2}). Nevertheless,
for the case of interest $m_b\approx 4.5$~GeV, neglecting the subleading
radiative corrections may provide a reasonable approximation for the endpoint
of the electron energy spectrum.

Before analyzing the general structure of the radiative corrections, it is
instructive to consider a simple example which illustrates the importance of
subleading radiative corrections.  Consider the order $\alpha_s$ correction
given in eq.~(\ref{radiative}).  This correction has $\ln^2(1-y)$ and
$\ln(1-y)$ singularities as $y\rightarrow 1$. The $\ln^2(1-y)$ singularity is
summed into the Sudakov form factor $R(y)$, leaving the subleading
$\ln(1-y)$  singularity. This subleading logarithmic singularity must also be
understood in order to determine the effect of radiative corrections on the
endpoint energy spectrum \cite{Politzer}. To see this, note that it is
possible to write two different expressions for the decay spectrum which
contain the same Sudakov leading singularity, but which have very different
behaviors as $y \rightarrow 1$.  The first expression is the conventional
definition~\cite{ACCMM}
\begin{equation}\label{IIIi}
     {d\Gamma_{\rm free}\over dy}=R(y){d\Gamma_0\over
     dy}\left[1-{2\alpha_s\over 3\pi}\widetilde G(y)\right]\,,
\end{equation}
where
\begin{equation}\label{IIIii}
     \widetilde G(y) = G(y)- \ln^2(1-y)\,.
\end{equation}
However, one can also rewrite the decay spectrum as
\begin{equation}\label{IIIiii}
     {d\Gamma_{\rm free}\over dy}={d\Gamma_0\over
     dy}\left[R(y)-{2\alpha_s\over 3\pi}\widetilde G(y)\right]\,,
\end{equation}
which is equally valid to order $\alpha_s$. The two expressions (\ref{IIIi})
and~(\ref{IIIii}) have the same $\ln^2(1-y)$ singularity as $y\rightarrow 1$,
but differ in the subleading terms. The first expression (\ref{IIIi}) vanishes
as $y\rightarrow 1$, whereas eq.~(\ref{IIIii}) diverges as $y\rightarrow 1$.
Thus, the exact form of the subleading singularity is required in order to
determine the shape of the spectrum very near the endpoint.

We will now demonstrate that in the limit $m_b\rightarrow\infty$, summing the
most singular $1/m_b$ corrections with eq.~(\ref{derivseries2}) cannot be used
to improve the behavior of the electron spectrum near $y=1$ without first
summing an infinite number of subleading perturbative QCD singularities.  In a
schematic notation in which we include only the powers of $\alpha_s$ and
$\ln(1-y)$, the radiative corrections near $y=1$ have the structure
\begin{eqnarray}\label{Rstructure}
        && 1\nonumber\\
        &+& \alpha_s\ln^2(1-y) + \alpha_s\ln(1-y) + \alpha_s\nonumber\\
        &+& \alpha_s^2\ln^4(1-y) + \alpha_s^2\ln^3(1-y) + \alpha_s^2\ln^2(1-y)
        + \alpha_s^2\ln(1-y) +\alpha_s^2\nonumber\\
        &+& \alpha_s^3\ln^6(1-y) +\alpha_s^3\ln^5(1-y) + \alpha_s^3\ln^4(1-y)
        + \alpha_s^3\ln^3(1-y) + \cdots\nonumber\\
        &+& \cdots\,.
\end{eqnarray}
The first column, containing terms of the form $\alpha_s^n\ln^{2n}(1-y)$,
exponentiates into the Sudakov factor $R(y)$, after which the most singular
terms remaining are of order $\alpha_s^n\ln^{2n-1}(1-y)$. We may write the
$m$th column of the expansion  (\ref{Rstructure}) as an infinite series of the
form
\begin{equation}\label{columnseries}
     C_m(y)=\sum_{n=[m/2]}^\infty b_{mn}\alpha_s^n\ln^{2n-m+1}(1-y)\,.
\end{equation}
The series of leading singularities corresponds to $m=1$; for this case, and
{\it only\/} this case, the coefficients
\begin{equation}\label{bno}
     b_{1n}={1\over n!}\left(-{2\over3\pi}\right)^n
\end{equation}
have been computed for all $n$, and the sum $C_1(y)$ is $R(y)$. The series
$C_m(y)$ for $m> 1$ represent an infinite set of infinite series, for which
the behavior of the coefficients $b_{mn}$ for large $n$ is not known.

The unknown subleading series $C_m(y)$, $m>1$, limit the accuracy with which
one can determine the electron energy spectrum. For perturbation theory to be
valid, one has to remain in a region in which all the subleading terms are
small, since their structure is not known, i.e.~all the terms beyond the first
column of eq.~(\ref{Rstructure}) must be small. This condition requires that
$\alpha_s^n\ln^{2n-m}(1-y) \ll 1$ for all $n$ and all $m> 1$, or that
$\alpha_s\ll1$ and
\begin{equation}\label{condition}
    \alpha_s \ln^2(1-y) < 1\,,
\end{equation}
which is the condition required for $n\rightarrow \infty$ with $m$ fixed. If
eq.~(\ref{condition}) is satisfied, the first column sums to $R(y)$, the
second  column is of order $\sqrt{\alpha_s}$ times the first column, the third
column is of order $\sqrt{\alpha_s}$ times the second column, and so on. The
condition (\ref{condition}) has converted the QCD perturbation series in
eq.~(\ref{Rstructure}) into an expansion in $\sqrt{\alpha_s}$. Summing all the
leading singularities $\alpha_s^n\ln^{2n}(1-y)$, or summing any finite number
of columns of eq.~(\ref{Rstructure}), does not increase the region of validity
of the perturbation expansion, since the condition that the next column be
small is still eq.~(\ref{condition}).  To increase the region of validity of
the perturbative expansion,  one must sum all the terms of the form
$\alpha_s^n \ln^{2n-m}(1-y)$ for $0\le m\le \lambda n$ and $\lambda>0$, in
which case one needs only the restriction $\alpha_s\ln^{2-\lambda}(1-y) < 1$.
That is, one must sum all the terms in (\ref{Rstructure}) below a line which
makes an angle $\tan^{-1} \lambda$ with the vertical, which implies that one
must sum a large number of subleading logarithms at high orders in perturbation
theory.

At present, only the sum $C_1(y)$ of the first column is known, so the
condition for the subleading QCD radiative corrections to be small is that
given by eq.~(\ref{condition}). To determine the restriction on $y$, we use
eq.~(\ref{condition}) in the form
\begin{equation}
     {\alpha_s(m_b)\over \pi}  \ln^2(1-y) < 1\,,
\end{equation}
where we have noted that the perturbation series is really in $\alpha_s/\pi$
rather than $\alpha_s$. The condition for reliability of the QCD radiative
corrections is
\begin{equation}\label{valid}
     1-y > e^{-\sqrt{\pi/\alpha_s}}\,.
\end{equation}
For very heavy quarks, this corresponds to a region that is much larger than
the smearing width $\epsilon$ of the $1/m_b$ corrections. To see this, take the
limit $m_b\rightarrow\infty$ with $\alpha_s$ at high energies held fixed,
i.e.~$\alpha_s(m_b)=6\pi/[(33-2n_f)\ln m_b/\Lambda_{\rm QCD}]$ ($n_f=5$).
Define the parameter $t$ by $\ln m_b/\Lambda_{\rm QCD}=t^2$.  Then
eq.~(\ref{valid}) becomes
\begin{equation}\label{validb}
     1-y > e^{-t \sqrt{23 / 6}}\,,
\end{equation}
whereas $\epsilon \sim e^{-t^2}$. For large quark masses (large $t$),
$\epsilon$ is much smaller than the restriction (\ref{validb}) on $1-y$.

As we have already noted, the residual momentum averaging procedure discussed
in Sect.~3 can be applied to the QCD corrected free quark decay spectrum. This
procedure yields the leading $1/m_b$ singularities to all orders in
$\alpha_s$.\footnote{The leading singularities arise when the derivatives in
eq.~(\ref{derivseries2}) act on the Sudakov suppression factor $R(y)$, not on
the $\theta$-functions as in eq.~(\ref{shapefunction}).}  When radiative
corrections are neglected, we have shown that the leading $1/m_b$ singularities
smear the decay spectrum by a width of order $\epsilon$.  Thus, a reliable
determination of the decay spectrum near the endpoint requires knowing the
lowest order (in $1/m_b$) spectrum at least within a distance $\epsilon$ of
the endpoint. Eq.~(\ref{valid}) implies that in the $m_b\rightarrow\infty$
limit, the lowest order spectrum with perturbative QCD corrections is not
known in a region near the endpoint which is much larger than $\epsilon$.

\section{Numerical Results and Conclusions}

The radiative corrections become large in a region given by eq.~(\ref{valid})
which is much larger than $\epsilon$ in the limit $m_b\rightarrow\infty$.
However, for a large but finite quark mass it is possible that one is in a
regime where $e^{-\sqrt{\pi/\alpha_s}}$ is not much larger than $\epsilon$.
For example, for $m_b=4.5$~GeV, we find $\alpha_s(m_b)\sim 0.2$ and
$e^{-\sqrt{\pi/\alpha_s}}\sim 0.02$. This value of $1-y$ corresponds to a
smearing width of approximately 50~MeV, which is smaller than $\epsilon$.
Whether this crude estimate is valid depends critically on the size of the
coefficients of the subleading terms in the second and higher columns of
eq.~(\ref{Rstructure}). For example, the subleading $\ln(1-y)$
and constant terms in $G(y)$ (see eq.~(\ref{Gdef})) give
\begin{equation}\label{subest}
     \left({2\alpha_s\over  3 \pi}\right)\left[\left({31\over 6}\right)\ln(1-y)
+\pi^2+{5\over4}\right]     \approx -0.1\,,
\end{equation}
when the electron energy is 200~MeV away from the free quark endpoint $y=1$.

Eq.~(\ref{subest}) indicates that it may be a good approximation to
include the effects of perturbative QCD corrections on the shape of the
endpoint region using for $d\Gamma_{\rm free}/dy$ the free b-quark decay
rate including the leading QCD double logarithms with
eq.~(\ref{exponentiated}). The smallness of eq.~(\ref{subest}) arises
from a cancellation between the $\ln(1-y)$ and constant terms in $G(y)$.
Each of these separately is not particularly small. Thus we
are not completely confident that higher order perturbative corrections are
negligible. It may be possible to sum the second column of
eq.~(\ref{Rstructure}) using the methods developed in \cite{comps}. Such a
summation would provide useful information on the importance of higher
order QCD corrections.

The shape of the endpoint region of the electron spectrum depends on the
matrix elements $\langle B(v)|\, iD^{\mu_1}\ldots iD^{\mu_n}\,|B(v)\rangle $.
Neubert estimates these matrix elements using a quark model for the $B$
meson~\cite{Neubert}. Eventually, these matrix elements can be determined
directly from experiment. For example, the same matrix elements occur in the
$1/m_b$ corrections to semileptonic $b\rightarrow c$ decay and in the decay
$b\rightarrow s\gamma$~\cite{self}. Thus a precise measurement of the electron
spectrum in $b\rightarrow c$ semileptonic decay can be used to obtain the
endpoint electron spectrum for $b\rightarrow u$ semileptonic decay and the
photon energy spectrum in $b\rightarrow s \gamma$.  The order $\alpha_s$
radiative corrections have also been computed for $b\rightarrow s\gamma$
\cite{AliGreub}. Let $x_\gamma = 2E_ \gamma/m_b$, and define
\begin{equation}
     F(y)=\int_{y}^1 {d\Gamma\over dx_\gamma} dx_\gamma\,,
\end{equation}
where $d\Gamma/dx_\gamma$ is the inclusive photon energy spectrum in
$b\rightarrow s\gamma$, neglecting the strange quark mass. $F(y)$ for
$b\rightarrow s\gamma$ and $d\Gamma/dy$ for $b\rightarrow u$ semileptonic
decays have the same $\ln^2(1-y)$ but different  $\ln(1-y)$
singularities as $y\rightarrow  1$. However, $F(y)$ for
$b\rightarrow s\gamma$ and $d\Gamma/dy$ for $c\rightarrow d$ semileptonic
decays do have the same $\ln^2(1-y)$ and $\ln(1-y)$
singularities as $y\rightarrow  1$ in the order $\alpha_s$ radiative
corrections.

The methods in this paper and ref.~\cite{Neubert} for describing the
endpoint region of the electron spectrum apply when the endpoint is
dominated by many states with masses of order $\sqrt{m_b\Lambda_{\rm
QCD}}$. However, in the non-relativistic constituent quark model
estimate of ref.~\cite{ISGW}, the region beyond the $B\rightarrow X_c e
\bar\nu_e$ endpoint is dominated by the single
decay mode $B\rightarrow \rho e \bar\nu_e$. If $\rho$ dominance is found
to hold experimentally, then the sum of the leading singularities is not
a valid description of the endpoint in a region which is as small as the
difference between the $B\rightarrow X_c e\bar \nu_e$ and $B\rightarrow
X_u e \bar\nu_e$ endpoints.

\acknowledgements
It is a pleasure to thank M.~Luke and H.D.~Politzer for helpful conversations.
This work was supported by the Department of Energy under Grants
Nos.~DOE-FG03-90ER40546 and DEAC-03-81ER40050, and by the Presidential Young
Investigator Program under Grant No.~PHY-8958081.

\end{document}